# Can Organic Ions Dissolve Fullerenes?

**Vitaly V. Chaban[1,], Cleiton Maciel[2], and Eudes Eterno Fileti[3]**

[1] MEMPHYS - Center for Biomembrane Physics, Odense M, 5230, Kingdom of Denmark

[2] Centro de Ciências Naturais e Humanas, Universidade Federal do ABC, 09210-270 Santo André, SP, Brazil

[3] Instituto de Ciência e Tecnologia, Universidade Federal de São Paulo, 12231-280, São José dos Campos, SP, Brazil

**Abstract**. Over 150 solvents have been so far probed to dissolve light fullerenes – with quite moderate success. Based on unusual electronic properties of the $C_{60}$ fullerene and room-temperature ionic liquids (RTILs), we hereby propose to use their mutual polarizability to significantly promote solubility of $C_{60}$. We report analysis supported by density functional theory with reliable hybrid functional and large-scale molecular dynamics simulations with specifically tuned empirical potential. The analysis suggests a workability of the proposed mechanism and provides qualitatively new insights into obtaining real fullerene solutions. Our preliminary findings open a wide avenue for engineering specific solvents for fullerenes and, possibly, some other nanoscale carbonaceous structures.

TOC image

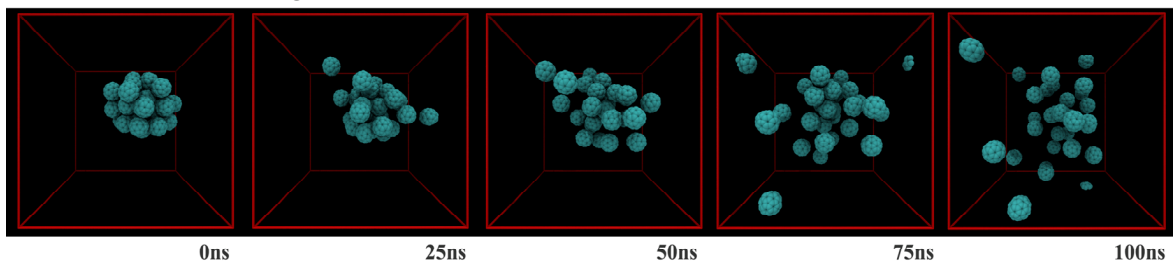

Efficient dissolution of fullerenes is an old quest started in 1985.[1] It is important for numerous applications including, but not limited to, development of cosmetic products, robust drug delivery, extraction and separation of nanostructures from industrial soot, photovoltaics, investigation of chemical reactions involving fullerenes, etc.[2-10] Solutions and colloidal systems of fullerenes belong to one of the most studied systems in modern chemistry. A variety of solvents (over 150) and their mixtures were applied to obtain real solutions of $C_{60}$, $C_{70}$, and some higher fullerenes.[4, 11, 12] The available data are, however, rather pessimistic. In the majority of polar solvents, including water, solutions of fullerenes are either extremely diluted or colloidal ones.[4] Relatively concentrated solutions are achieved in some solvents of low polarity, such as aromatic hydrocarbons and their haloderivatives. For instance, solubility of $C_{60}$ in benzene equals to 1.50 g l$^{-1}$,[13] 2.40 g l$^{-1}$ in toluene,[14] 2.60 g l$^{-1}$ in ethylbenzene,[13] all measurements being done at 298 K. Remarkably, methylation of the solvent molecule significantly increases its affinity to $C_{60}$, although chemical background of this empirical finding is unclear. 1,2,3-trimethylbenzene allows for solubility increase by more than three times, compared to benzene, up to 4.70 g l$^{-1}$.[15] The same authors reported 17.90 g l$^{-1}$ for 1,2,4-trimethylbenzene and 5.80 g l$^{-1}$ for 1,2,3,4-tetramethylbenzene,[15] although the first quantity looks doubtful for us. Halogenation of low molecular aromatic compounds provides even better results in many cases: 6.35 in chlorobenzene, 13.80 in 1,2-dibromobenzene, 9.62 in 1,2,4-trichlorobenzene, etc.[12] Grafting other functional groups does not lead to comparable effects. Benzonitrile, nitrobenzene, phenyl isocyanate, n-propylbenzene, *n*-butylbenzene and a few other solvents, carefully summarized by Semenov et al.,[11, 12] are notably unsuccessful for pristine light fullerenes.

The highest solubility of $C_{60}$ that we are aware of was achieved by Talukdar and coworkers,[14] who used piperidine (53.28 g l$^{-1}$) and pyrrolidine (47.52 g l$^{-1}$). It was hypothesized that these solvents exhibited specific donor-acceptor interactions with $C_{60}$, $C_{70}$, $C_{84}$, $C_{100}$, that can, under certain conditions, give rise to formation of new chemical compounds. Piperidine and pyrrolidine are sometimes called "reactive" solvents in respect to $C_{60}$. Note, that published references on solubility contain lots of contradicting results. This fact can be correlated both with specific methods of measurements and their internal uncertainties and preparation of fullerene containing solutions.[4, 12, 16-18] In all polar and many low-polarity solvents, fullerenes tend to form colloidal systems, being another source of the scattered, ambiguous results.

Apart from direct experiments outlined above, numerous computer simulations have been conducted on fullerenes in solutions.[19-22] Monticelli and coworkers published a few important developments in atomistic and coarse-grained force fields,[19, 23, 24] one of us reported free energies and structure patterns of $C_{60}$ in various solvents,[25-27] a few high-impact works were devoted to the partitioning of $C_{60}$ and its derivatives at the water/bilayer interface.[21, 28] Density functional theory (DFT) was applied to probe applicability of fullerenes in photovoltaics and obtaining fundamental understanding of their electronic structures.[10, 29-32] In the meantime, we are not aware of any investigations of the fullerene dissolution process in real time and length scales.

To summarize, even if "reactive" solvents are counted, pristine fullerenes cannot currently be solvated in the same sense as conventional, smaller molecules can be. Unlike oxygens in ozone, carbons in fullerenes are chemically equivalent and, therefore, no permanent electric moment exists. Short-range dispersive attraction in case of carbon is inferior to that of more electronegative elements, but is strong enough to keep fullerenes in a solid state at ambient conditions. Because of the fullerene's supramolecular size, its introduction into solvent makes entropic part of solvation free energy, $T dS$, significantly negative, whereas enthalpic constituent is never high enough. The time has come to look at solvation of fullerenes from a new molecular perspective.

In the present work, we introduce a novel concept of solvating fullerenes using room-temperature ionic liquids (RTILs). Our concept utilizes genuine high electronic polarizability of fullerenes, as compared to other carbonaceous compounds, and peculiarities of delocalized electron density in organic cations of RTILs. The original evidence of our method efficiency was obtained from functional density functional theory coupled with highly accurate hybrid exchange-correlation functional, and large-scale atomistic-precision molecular dynamics (MD) simulations.

*Insights into non-bonded fullerene-ionic liquid interactions*. In their condensed state, most ionic liquids[33-35] are very polarizable and there is an electron density transfer between anion and cation.[36-38] We selected one of such RTILs, 1-butyl-3-methylimidazolium tetrafluoroborate, [BMIM][BF$_4$], to demonstrate our new concept. In turn, fullerenes are also known for high dipole polarizability, 80 Å$^3$,[39] because of unusual electronic structure and icosahedral symmetry. In order to measure the extent of nonadditivity of C$_{60}$-RTIL non-bonded interactions, we employed density functional theory (DFT) methodology. We described an electron density using a high-quality hybrid exchange-correlation functional, omega B97X-D.[40] Within this functional, exchange energy is combined with the exact energy from Hartree-Fock theory. Furthermore, empirical atom-atom dispersion correction is included, being of vital importance in case of fullerenes. Hybrid functionals have been shown to bring significantly improved accuracy in the resulting electronic structure of the simulated quantum mechanical systems, and therefore, polarization effects. While many developments, in particular based on local density approximation, tend to overestimate electron transfer between nuclei, carefully "trained" hybrid functionals are free of this defect. According to Chai and Head-Gordon, the functional simultaneously yields satisfactory accuracy for thermochemistry, kinetics, and non-covalent interactions. Based on numerous tests, omega B97X-D is superior for non-bonded interactions over previous hybrid functionals.[40] The 6-31G(d) basis set, containing vacant *d*-orbitals, was used in all computations, since this basis set is continuously applied to describe a great variety of carbonaceous compounds. Whereas 6-31G(d) is less comprehensive than Dunning's correlation-consistent basis sets (especially triple, quadruple, and higher order implementations[41]), it provides reasonable accuracy in case of ground-state systems. Note that implementation of

6-31G(d) instead of aug-cc-PVTZ[41] decreases computational cost of the same DFT method by at least one order of magnitude.

The system containing one $C_{60}$ and 500 [BMIM][BF$_4$] was originally simulated using empirical scaled-charge potential for [BMIM][BF$_4$], introduced by one of us earlier.[36] The $C_{60}$ fullerene was represented by suitable AMBER-type Lennard-Jones (12,6), LJ, carbon parameters.[42] The set of empirical parameters for $C_{60}$ and [BMIM][BF$_4$] is listed in Ref.[43] After equilibrium had been reached at 300 K, we extracted Cartesian coordinates of $C_{60}$ with its solvation shells. Solvation shells within the radius of 0.8 (5 ion pairs), 1.0 (10 ion pairs), 1.2 (24 ion pairs), and 1.4 nm (34 ion pairs), Figure 1, were generated. $C_{60}$ with just one ion pair nearby was used for comparison. Binding energy, corresponding to non-bonded interactions in each shell, electron localization, and full orbital structure were calculated using omega B97X-D/6-31G(d). Basis set superposition error (ca. 20% of average binding energy in our case) was corrected using the counterpoise technique. Figure 2 suggests that certain electron transfer indeed occurs between ions and $C_{60}$. Brought into direct non-bonded contact with RTIL, fullerene lost certain part of electron density, resulting in the +0.35e charge. The depicted charges were obtained using real-space integration of electron density within certain radius from the nucleus, based on the classical Hirshfeld technique. The total electron charge on $C_{60}$ stabilizes after number of ions achieves the size of the fullerene's first solvation shell. Additional increase of the shell size does not lead to any alteration of the electron transfer phenomenon.

The binding energies (Figure 3) between $C_{60}$ and RTIL suggested by original potential (essentially based on the Lorenz-Berthelot prediction) appear underestimated by 15-20%. We correlate this underestimation with the electron transfer discussed above (Figure 2). The transfer is promoted via disturbance of the fullerene's highest energy electrons by electric field of neighboring ions, although a complete ionization of $C_{60}$ does not take place in these systems. Since such kind of non-additive interaction is caused by induced dipole, it should be described by a specific energetic well depth parameter, epsilon, in LJ potential, rather than adjusting permanent Coulombic charges on the $C_{60}$ and ions. Iteratively changing epsilons for pairwise $C_{60}$-BMIM$^+$ and $C_{60}$-BF$_4^-$ interactions, we found an optimal set of parameters to approximate non-bonded energetics predicted by DFT (Figure 3). Note, that well depth parameters were changed uniformly, irrespective of nature of each chemical element. Pairwise epsilons were changed only for polar moieties of RTIL. The workability of this approximation to accurately describe (Figure 3) binding in case of each of the five shells is a key evidence in favor of our methodological solution. After the refinement of the force field, the MD system was additionally equilibrated, new solvation shells were produced, and DFT calculations were repeated. However, the resulting binding energies were nearly the same as in the first case. Therefore, solvation shell structures and ion orientations near the fullerene surface remained intact. According to snapshots in Figure 1, $C_{60}$ was preferentially coordinated by cations. Interestingly, cations coordinated $C_{60}$ by their polar moiety, imidazole ring, and not by the hydrophobic -CH$_2$-CH$_2$-CH$_2$-CH$_3$ chain. The qualitative observations were the same, irrespective of using original or refined FFs. Such orientation is, likely, more preferable for periodic system due to relatively strong dispersion-type attraction between nitrogen atoms of imidazole ring and carbon atoms of fullerene.

Although RTIL imposes certain polarization on $C_{60}$, no *chemical bonding* occurs between these particles. In case of the first solvation shell (10 ions pairs) the binding energy is 30 kJ/mol at 300 K per [BMIM][BF$_4$] ion pair. Being normalized per number of atoms in the polar part of cation and anion, it is much weaker than conventional hydrogen bonding. A few highest energy π-orbitals of $C_{60}$ are responsible for the observed electron transfer. Figures 2 and 3 predict that 0.1 to 0.35e are pulled from fullerene to anion (i.e. to more electronegative fluorine atoms). Interestingly, the second solvation shell (see values corresponding to 24 and 34 ion pairs) does not contribute to electron transfer, although number of ions in the first solvation shell plays a major role (see values corresponding to one, five, and ten ions pairs). In future investigations, it would be instructive to gain insights about charge transfer dynamics and extend studies to another temperature range, since electronic polarization is known to depend on temperature.

Remarkably, solvation leads to energy increase by 0.1-0.3 eV (depending on the shell size) of highest occupied and lowest unoccupied orbitals, localized on fullerene. Such significant shift normally indicates a strong coupling between the solute and the solvent.

*Insights from large-scale molecular dynamics.* A solid species of the $C_{60}$ fullerene was immersed into 1-butyl-3-methylimidazolium tetrafluoroborate (1500 ion pairs), Figure 4. Non-equilibrium molecular dynamics was simulated at 300, 310, 320, 350, and 400 K under ambient pressure, until the systems came to thermodynamic equilibrium. Equilibrium properties were derived using 30 ns long trajectories with a time-step of 0.002 ps and coordinates saving frequency of 20,000 frames per nanosecond. Velocity rescaling thermostat[44] and Parrinello-Rahman barostat[45] with coupling constants of 0.1 ps and 4.0 bar were turned on to represent NPT ensembles of these systems. Covalent bonds of heavy elements with hydrogen atom were constrained using the LINCS algorithm[46], which allows for integration time-step increase without influencing computational stability. The force field was refined as described above. Note, that our model provides average $C_{60}$-RTIL and RTIL-RTIL energetics ideally concordant with average energetics from density functional theory method. The Lennard–Jones interactions were shifted to zero between 1.2 and 1.3 nm. The real-space Coulomb interactions were truncated at 1.4 nm. Their long-range parts were taken into account via the particle-mesh Ewald method.[47] All large-scale simulations were carried out using the GROMACS 4 software package.[48-51]

Indeed, fullerene is nearly insoluble at 300 K. Although the well-ordered periodic structure of the solid phase changes to cluster-like (Figure 4), no fullerene is present in dissolved state (Figure 5) for a significant time during 300 ns of the spontaneous evolution of MD system. This result is in excellent agreement with recent experimental observations,[52] where a few popular RTILs were probed for the $C_{70}$ solvation at room temperature. None of the RTILs, such as 1-butyl-2,3-dimethylimidazolium bis(trifluoromethylsulfonyl)imide, 1-methyl-3-

octylimidazolium tetrafluoroborate, 1-methyl-3-octylimidazolium hexafluorophosphate, 1-methyl-3-octylimidazolium bis(trifluoromethylsulfonyl)imide, 1-decyl-3-methylimidazolium tetrafluoroborate, methyltrioctylammonium bis(trifluoromethylsulfonyl)imide, trihexyltetradecylphosphonium chloride, etc, appeared an efficient solvent for fullerene, according to fluorescent measurements reported by Martins et al.[53] However, some of the above RTILs demonstrated better solvation of $C_{70}$ as compared to [BMIM][BF$_4$], from 0.02 up to 0.06 g l$^{-1}$. Unfortunately, all cites values are nearly negligible.

Despite failure at ambient conditions, [BMIM][BF$_4$] becomes outstandingly successful upon just 20 K temperature increase, 5 g l$^{-1}$ (310 K), 49 g l$^{-1}$ (320 K), > 66 g l$^{-1}$ (333 K). Since enthalpic factor does not change significantly between 300 and 320 K[9], the dissolution is boosted due to entropic factor increase. We were unable to identify *maximum* $C_{60}$ solubility at higher temperatures, since the originally scheduled systems contained only 30 fullerene molecules corresponding to a maximum solubility of 66 g l$^{-1}$. This limit was achieved and exceeded at 333 K. Compare 66 g l$^{-1}$ to the solubility potential of "reactive" solvents,[14] although no reaction between $C_{60}$ and RTIL is anticipated in our case.

The above findings look very encouraging and motivate to take a closer look at the representatives of other families of RTILs in order to foster solubility towards ever higher values. Mutual polarization of [BMIM][BF$_4$] and $C_{60}$ plays a key role in successful dissolution. Ionic nature of the [BMIM][BF$_4$] liquid and delocalized charge on the cation are important prerequisites for their electronic polarizability. Note, that our approach is possible only with RTILs, but impossible with solutions of ions in *polar* molecular liquids. In the latter case, ions are strongly solvated by the solvent molecules and cannot approach the fullerene. The fullerene is surrounded predominantly by neutral particles, which, in most cases, exhibit insignificant (e.g. alcohols) to negligible (e.g. nitriles) polarization potential.

Figure 5 depicts radial distribution functions between fullerene center-of-masses. At 300K, undissolved cluster (Figure 4) is evidenced by sharp, well-ordered peaks. The arrangement of peaks in not ideal though, as it is observed at lower temperatures, below 260 K (not shown). Therefore, [BMIM][BF$_4$] exhibits a strong potential for breaking the C$_{60}$-C$_{60}$ non-chemical bonds even at ambient conditions, and insignificant changes of entropic or enthalpic factors can alter solubility drastically. This readily happens at higher temperatures (e.g. 320-333 K), resulting in the unexpectedly high value of minimum 66 g l$^{-1}$. Larger clusters of C$_{60}$ must be simulated at T > 333 K to estimate and tabulate solubility. On a related note, Li and co-workers have recently conducted Raman and IR measurements, as well as DFT calculations, to investigate the dispersion of single-walled carbon nanotubes in imidazolium ionic liquids at room temperature.[53] On the contrary to their expectations, all methods unequivocally suggested no strong interaction, such as π- π, between carbon nanotubes and imidazolium cations. The only recorded interactions were weak van der Waals attraction, indicating that RTILs are ideal solvents to disperse nanotubes without influencing its electronic structure and properties.

The distance of 0.8 nm defines the first coordination sphere of the cluster at 300 K (Figure 5, left), i.e. neighboring fullerene molecules are aggregated, as the distance between their centers does not exceed 0.8 nm. With this information in mind, it was possible to analyze the time evolution of the dissolution process in terms of the maximum number of molecules in the cluster (Figure 5, right). The dissolution process takes ca. 20 ns at 400 K and ca. 100 ns at 350 K. Note, however, that dissolution time/speed depends on the volume of solid phase and on interface surface. With the above data from MD simulations, speed of C$_{60}$ solvation can be assessed for various aggregate sizes and various external conditions. Statistics of the individual clusters size (Figure 6), accumulated during an equilibrium stage of simulation, clearly illustrates that dissolution takes place, and this process is unambiguously thermodynamically favorable. Interestingly, a few dimers and trimers are possible in the C$_{60}$ solutions in [BMIM][BF$_4$], although their fractions do not exceed 10%. Such dimers and trimers are wide-spread in the aqueous and nonaqueous electrolyte solutions.

**Conclusion**

To recapitulate, a new approach to solvate $C_{60}$ was outlined in the present work. Although a zero solubility in [BMIM][BF$_4$] was recorded at 300 K (in perfect agreement with experimental observations[52]), the solubility increases drastically with temperature: S (310 K)=5 g l$^{-1}$, S (320 K)=49 g l$^{-1}$, S (333 K)>66 g l$^{-1}$. The accuracy of the results is justified by carefully parameterized non-bonded interactions among all components in the simulated system, including electron structure calculations. Although our atomistic-precision MD experiments cannot compete with all-electron description, it is the only computationally feasible approach to derive insights for the appropriate time (over 300 ns) and length (over 500 nm$^3$) scales.

Here, we chose [BMIM][BF$_4$] as an exemplary RTIL due to its wide availability and usability in fundamental research and technology.[36, 38, 52, 54] We suppose that certain other RTILs can exhibit even higher affinity towards fullerene (enthalpic factor). As fullerene exhibits π-stacking with many other aromatic compounds, one can hypothesize that polyaromatic cations, such as isoquinolinium$^+$, isothiouronium$^+$, triphenylphosphonium$^+$, may appear additional decent solvents. In the latter case, both electronic polarization and π-stacking are involved in the intermolecular binding. Unfortunately, we currently do not have trustworthy information about liquid ranges of these families of RTILs.

Trace admixtures of salts containing alkali ions, such as Li$^+$, Na$^+$, K$^+$, may improve solubility, provided that small, mobile cations localize near the $C_{60}$ surface or penetrate inside the cavity. Previous study suggested that even a single lithium atom inside $C_{60}$ greatly perturbs electronic energy levels of the whole structure.[10]

Experimental studies on $C_{60}$ solubility in [BMIM][BF$_4$] and other mentioned RTILs upon a wide temperature range are urgently requested.

**Acknowledgements**

The computations have been partially supported by the Danish Center for Scientific Computing (Horseshoe 5). C. M. and E. E. F. thank Brazilian agencies FAPESP and CNPq for support.

**Authors for Correspondence**

V. V. C.: Tel. 45-6550-4040. Email: vvchaban@gmail.com; chaban@sdu.dk

E. E. F.: Tel. 55-12-3309-9500. Email: fileti@gmail.com; fileti@unifesp.br

**FIGURE CAPTIONS**

Figure 1. The first solvation shells of fullerene. All ions of the shell are within 0.8 (a), 1.0 (b), 1.2 (c), and 1.4 nm (d) from $C_{60}$.

Figure 2. Electron charge on cation, anion, and fullerene as a function of ion pairs number.

Figure 3. Binding energy between $C_{60}$ and [BMIM][BF$_4$] as a function of ion pairs number: (a) normalized per mole of fullerenes; (b) normalized per mole of [BMIM][BF$_4$]; (c) percentage difference between original and refined models. Original model is red circles, DFT energy is green circles, refined model is blue triangles.

Figure 4. Representative configuration snapshots of the solid (left) and dissolved (right) state of $C_{60}$.

Figure 5: (Left) radial distribution function of the center of mass of fullerenes; (right) evolution of the largest cluster size versus simulation time.

Figure 6: Monomers, dimers, and trimers of $C_{60}$ in [BMIM][BF$_4$].

FIGURE 1

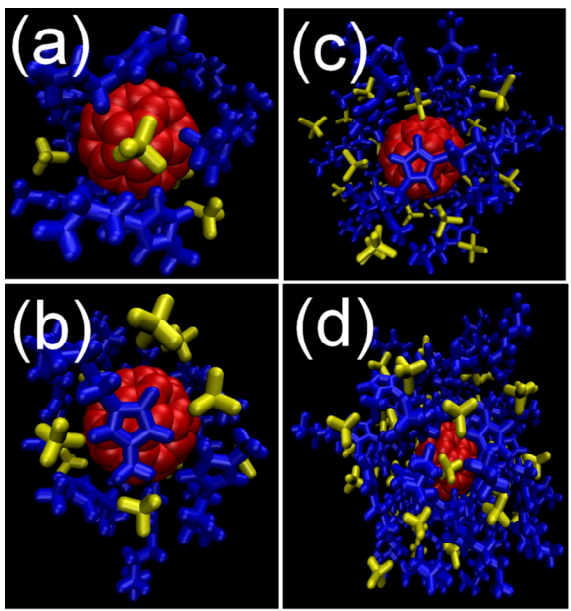

FIGURE 2

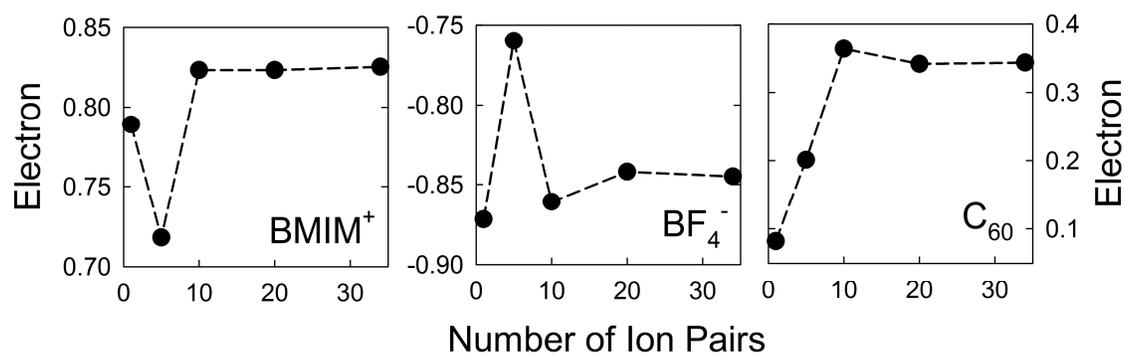

FIGURE 3

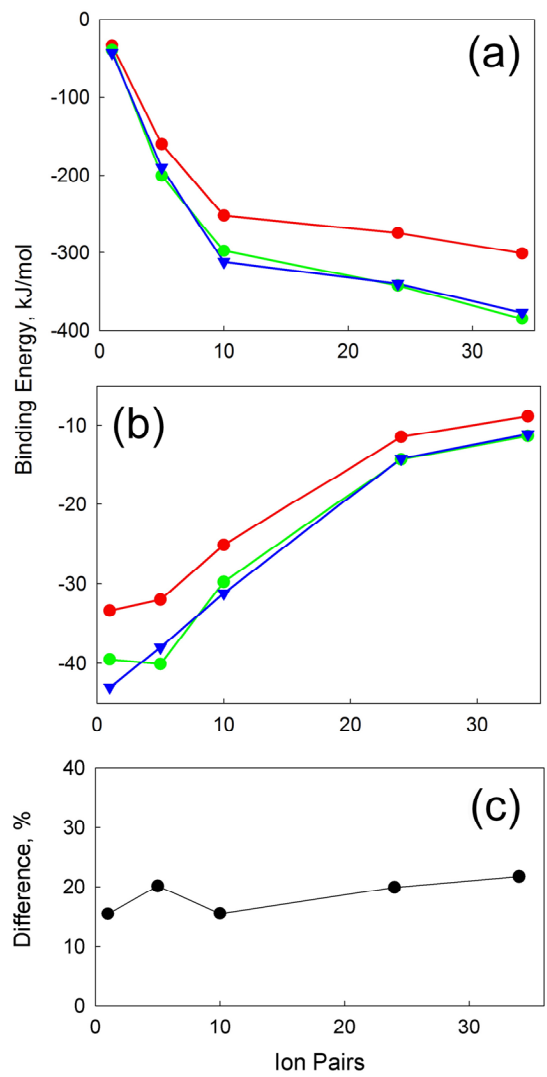

FIGURE 4

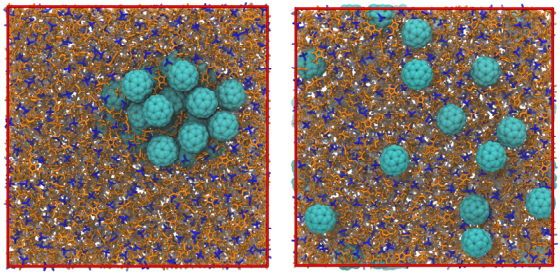

FIGURE 5

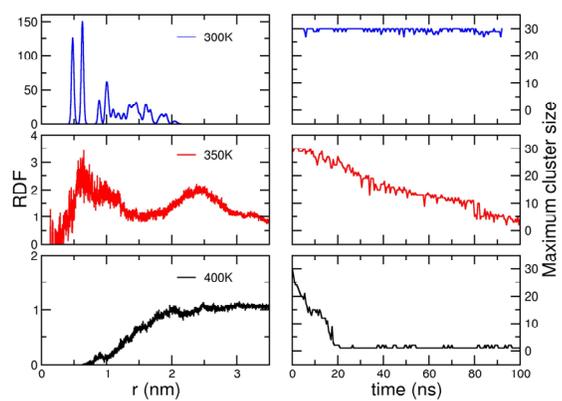

FIGURE 6

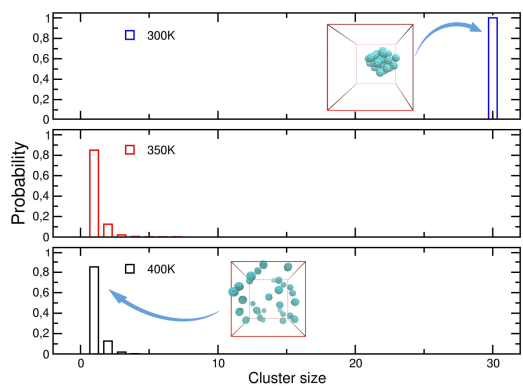